\documentclass[%
 reprint,
superscriptaddress,
 amsmath,amssymb,
 aps,
]{revtex4-1}

\usepackage{graphicx}
\usepackage{dcolumn}
\usepackage{bm}

\usepackage{gensymb}	
\usepackage{epsfig}		
\usepackage{multirow} 	

\begin{document}

\preprint{APS/123-QED}

\title{Infrared study of Large scale h-BN film and Graphene/h-BN heterostructure}

\author{Kwangnam Yu}
\affiliation{Department of Physics, University of Seoul, Seoul 130-743, Republic of Korea}

\author{Jiho Kim}
\affiliation{Department of Physics, University of Seoul, Seoul 130-743, Republic of Korea}

\author{Chul Lee}
\affiliation{Department of Physics, University of Seoul, Seoul 130-743, Republic of Korea}

\author{A-Rang Jang}
\affiliation{Department of Chemistry, Ulsan National Institute of Science and Technology (UNIST), Ulsan 689-798, Republic of Korea}

\author{Hyeon Suk Shin}
\affiliation{Department of Chemistry, Ulsan National Institute of Science and Technology (UNIST), Ulsan 689-798, Republic of Korea}

\author{Keun Soo Kim}
\affiliation{Department of Physics and Graphene Research Institute, Sejong University, Seoul 143-747, Republic of Korea}

\author{Young-Jun Yu}
\affiliation{Creative Research Center for Graphene Electronics, Electronics and Telecommunications Research Institute (ETRI), Daejeon 305-700, Korea}

\author{E. J. Choi}
\email[Corresponding author: ]{echoi@uos.ac.kr}
\affiliation{Department of Physics, University of Seoul, Seoul 130-743, Republic of Korea}

\date{\today}

\begin{abstract}
We synthesize a series of CVD h-BN films and perform critical infrared spectroscopic characterization. 
For high-temperature (HT, Temp~=~1400~$\celsius$) grown h-BN thin film only $E_{\rm{1u}}$-mode infrared phonon is activated demonstrating highly aligned 2D h-BN planes over large area, 
whereas low-temperature (LT, Temp~=~$1000$~$\celsius$) grown film shows two phonon peaks, $E_{\rm{1u}}$ and $A_{\rm{2u}}$, due to stacking of h-BN plane at tilted angle.
For CVD graphene transferred on HT h-BN/SiO$_{2}$/Si substrate, interband transition spectrum $\sigma_{1}$ shifts strongly to lower energy compared with that on LT h-BN/SiO$_{2}$/Si and on bare SiO$_{2}$/Si substrate, 
revealing that residual carrier density $n$ in graphene is suppressed by use of HT h-BN layer. 
Also the interband transition width of $\sigma_{1}$ defined by effective temperature is reduced from 
400~K for G/SiO$_{2}$/Si to 300~K for HT h-BN/SiO$_{2}$/Si. The behaviors of $n$ and effective temperature 
show that HT h-BN film can decouple CVD graphene from the impurity and defect of SiO$_{2}$ leading to large scale free-standing like graphene.
\end{abstract}

\maketitle

\section{Introduction}

Hexagonal boron nitride (h-BN) is a graphite-like layered material characterized by wide band gap $\sim$~6~eV.\cite{watanabe2004direct}
The B and N atoms bond covalently to constitute BN honeycomb planes, whereas adjacent BN planes are weakly coupled by van der Waals interaction.\cite{geim2013van}
h-BN shows a range of attractive properties such as high-temperature stability, high mechanical strength, and large thermal conductivity useful for application.\cite{sichel1976heat,watanabe2009far,sugino2000dielectric,jo2013thermal}
In particular when used as substrate for graphene h-BN greatly enhances carrier mobility at room- and low-temperature compared with conventional SiO$_{2}$/Si substrate.\cite{dean2010boron,gannett2011boron,kim2011chemical}
The superior performance as substrate is directly associated with the ultraflat, chemically inert surface and also the high purity, defect-free bulk nature of h-BN.\cite{lee2011electron,xue2011scanning}
However, the exfoliated h-BN flake is limited in size by small lateral length, only $\sim10\mu m$~$\times$~$10\mu m$, which seriously hinders its use in application. 
For practical industrial application large scale h-BN is strongly needed. 

Recently large area h-BN thin films are synthesized by CVD technique.\cite{kim2011synthesis,gao2013repeated,kim2013growth,park2014large,ismach2012toward}
Here chemical precursors containing B and N are thermally evaporated and deposited on metallic platforms (Cu, Pt, Fe)\cite{kim2011synthesis,gao2013repeated,kim2013growth,park2014large} or $\alpha$-Al$_{2}$O$_{3}$ crystal.\cite{ismach2012toward} 
Hexagonal phase BN film  is formed for certain ranges of vapor pressure and growth temperature.
The h-BN film is isolated from its template and transfered onto desired substrate such as SiO$_{2}$/Si.
Further transfer of a pregrown CVD graphene sheet on h-BN/substrate completes the large scale fabrication of graphene supported by h-BN buffer layer.\cite{suk2011transfer}
A systematic spectroscopic study and critical comparison with the exfoliated samples, currently lacking, can lead to deeper understanding of the large area samples.

In this work we perform infrared transmission measurement on the CVD h-BN and graphene/h-BN films.  
For the h-BN film infrared-active phonon peaks are probed. 
A correlation of the phonon profile with the growth condition is found, providing a knowledge about the lattice structure of the films.    
For graphene/h-BN interband transition of Dirac carrier in graphene is measured using the infrared spectroscopy.
From the electronic transition we are able to determine Fermi energy, carrier density, and defect-related interband transition width of graphene.

\section{Experiment}
We grew multilayer h-BN thin films using CVD method on Al$_{2}$O$_{3}$ substrate at two different temperatures (Temp) $1000$~$\celsius$ and $1400$~$\celsius$.  
The details of the film synthesis were reported elsewhere.\cite{jang2016wafer}
The h-BN films were isolated from the Al$_{2}$O$_{3}$ and transferred on IR-transparent SiO$_{2}$/p-Si substrate, 1~cm~$\times$~1~cm in size, by using electrochemical bubbling-based method.\cite{kim2013growth,jang2016wafer}
We also prepared single crystal multilayer h-BN flakes by exfoliation. 
Large scale monolayer graphene was grown on Cu-foil (Alfa Aesar, 99.999\%) using CVD method.\cite{li2009large,bae2010roll}
The CVD graphene was then transfered on the CVD h-BN/SiO$_{2}$/p-Si following the PMMA transfer method.\cite{suk2011transfer}
Infrared transmission was measured using a microscope FTIR spectrometer (Bruker Vertex 70V). 
Optical dielectric functions were calculated by fitting transmission and reflection data using RefFit analysis program.\cite{kuzmenko2005kramers}

\section{Result}

\subsection{h-BN thin film}

\begin{figure}[t!]
\centering
\epsfig{file=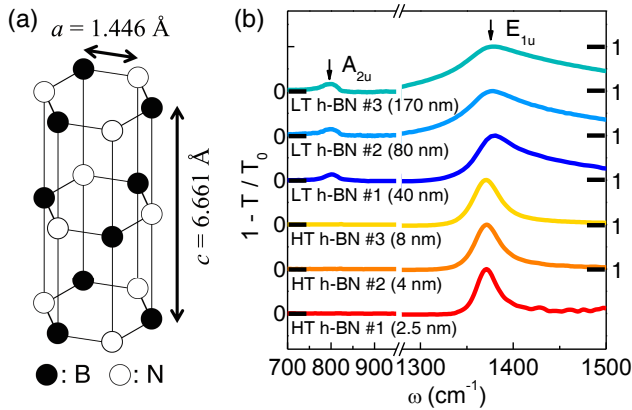,width=0.47\textwidth,angle=0,clip=}
\caption{
(a) Lattice structure of h-BN: space group = P6$_{3}$/mmc, lattice constants $a=1.446\AA$, $c=6.661\AA$.\cite{sichel1976heat}
(b) $1-T/T_{0}$ of h-BN film grown at low-temperature (LT, Temp~=~1000~$\celsius$) and high-temperature (HT, Temp~=~1400~$\celsius$). 
$T$ and $T_{0}$ stand for infrared transmission of h-BN/substrate and bare substrate respectively.
For LT h-BN, $A_{\rm{2u}}$ phonon is observed at $\omega=800$~cm$^{-1}$ which is absent for the HT h-BN film.
For each spectrum the intensity is normalized by that of $E_{\rm{1u}}$ peak. 
The numbers in the parenthesis show the thicknesses of the h-BN films.
}
\label{fig:fig.1}
\end{figure}

Figure 1(b) shows $1-T/T_{0}$ of the h-BN films grown at low-temperature (LT, Temp~=~1000~$\celsius$) and high-temperature (HT, Temp~=~1400~$\celsius$).
For LT h-BN two IR-active phonon peaks are observed at $\omega=800$~cm$^{-1}$ and $\omega=1375 $~cm$^{-1}$. 
On the other hand, for HT h-BN, the low-frequency peak is absent and only the high-frequency $E_{\rm{1u}}$ peak is activated.
The $E_{\rm{1u}}$ phonon represents in-plane oscillation of B and N atoms within the BN plane. 
The $\omega=800$~cm$^{-1}$ phonon is considered to be a $c$-axis mode associated with vibration normal to the BN plane.\cite{geick1966normal,hoffman1984optical} 
For our near-normal incident experiment, infrared light should couple only with the in-plane $E_{\rm{1u}}$ phonon mode if the BN-plane of the thin film were perfectly aligned with the substrate-plane.
The phonon profiles of LT h-BN and HT h-BN suggest that the two kinds of films differ in their lattice alignments.

\begin{figure}[t!]
\centering
\epsfig{file=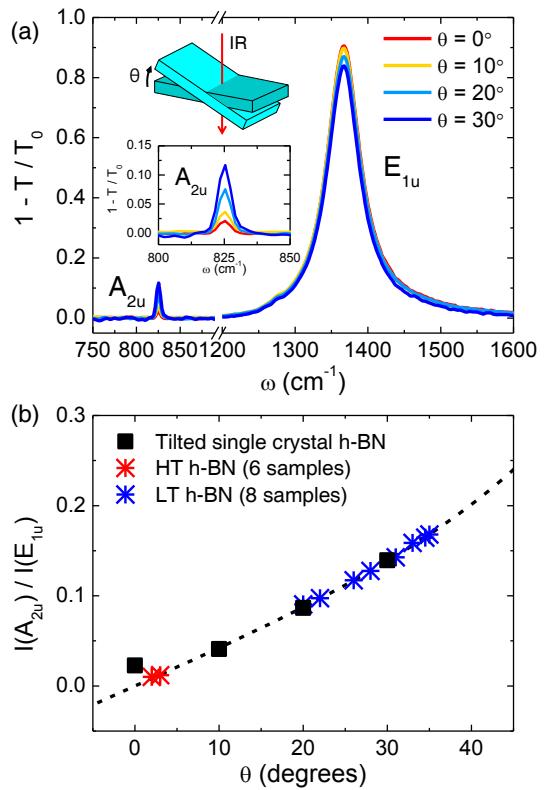,width=0.40\textwidth,angle=0,clip=}
\caption{
(a)
$1-T/T_{0}$ measured for a single crystal h-BN flake.
As the tilting angle $\theta$ increases, the $A_{\rm{2u}}$ phonon intensity increases 
(see the inset) and $I(E_{\rm{1u}})$ phonon intensity decreases respectively.
(b) 
Ratio of peak heights $I(A_{\rm{2u}})$ and $I(E_{\rm{1u}})$ as function of $\theta$.
The dashed curve, $0.24\tan\theta$, connects the single crystal data points. 
Based on this curve we infer the tilting angle of LT h-BN films $\theta=20\degree\sim40\degree$ from their phonon intensity ratio $0.1<I(A_{\rm{2u}})/I(E_{\rm{1u}})<0.15$.  
For HT h-BN films we have $I(A_{\rm{2u}})/I(E_{\rm{1u}})<0.01$ and $\theta<3\degree$.
}
\label{fig:fig.2}
\end{figure}

To better understand the phonon behavior we performed angle-tilted IR transmission measurement using single crystal h-BN.
Here 100~nm-thick h-BN crystal is tilted from $\theta=0$ to $30\degree$ with respect to the incident IR as sketched in Figure 2(a). 
As $\theta$ increase $E_{\rm{1u}}$ peak intensity decrease and that of $A_{\rm{2u}}$ peak increases.
The peak intensity $I(A_{\rm{2u}})$ and $I(E_{\rm{1u}})$ change with $\theta$ as $I(A_{\rm{2u}})\sim\sin\theta$ and $I(E_{\rm{1u}})\sim\cos\theta$ (data not shown). 
The $\theta$-dependent behaviors of the IR phonons confirm that they are in-plane and out-of-plane vibration modes respectively.
In Figure 2(b) we plot the intensity ratio $I(A_{\rm{2u}})/I(E_{\rm{1u}})$ as function of $\theta$. 
The ratio increases in proportion to tan$\theta$, $I(A_{\rm{2u}})/I(E_{\rm{1u}})=A\tan\theta$ ($A=0.24$), as shown by the dashed curve.
We observe very weak, non-zero $A_{\rm{2u}}$ peak at $\theta=0\degree$ because in our microscope spectrometer the IR light is focused on the sample with near-normal incidence angle. 
For LT h-BN the phonon intensity of the samples, eight in total, is confined in $0.1<I(A_{\rm{2u}})/I(E_{\rm{1u}})<0.15$. 
In terms of $\theta$ it corresponds to the tilting of the BN plane by $\theta=20\degree\sim40\degree$ according to the $I(A_{\rm{2u}})/I(E_{\rm{1u}})$-curve map.
In contrast all HT h-BN samples, six in total, show $I(A_{\rm{2u}})/I(E_{\rm{1u}})<0.01$ and therefore small tilting of the BN plane, $\theta<3\degree$.
For the film synthesis at Temp~=~1000~$\celsius$ (LT) our result shows that although the B and N precursors do form the hexagonal BN phase in CVD chamber, the BN planes fail to conform to the [0001] plane of the Al$_{2}$O$_{3}$ substrate. 
Instead tilted growth of BN plane is energetically favored. 
For high-temperature Temp~=~1400~$\celsius$ growth the tilting problem is largely cured, leading to almost perfectly aligned BN planes over the entire area of the large h-BN/substrate sample.

\begin{figure}[t!]
\centering
\epsfig{file=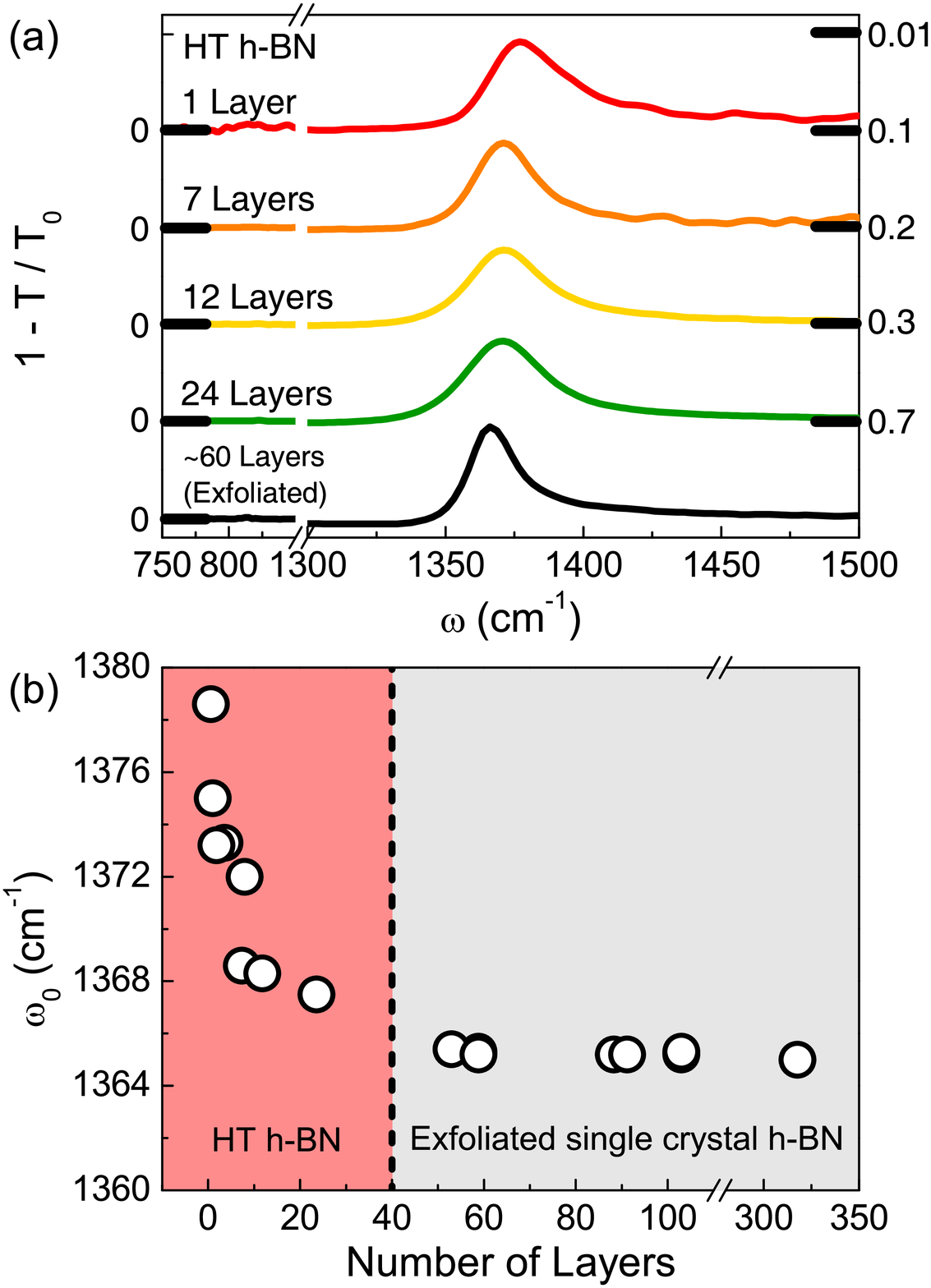,width=0.40\textwidth,angle=0,clip=}
\caption{
(a) 
Infrared phonon of HT h-BN with the film thickness ranging from 24 layers to 1 layer. 
For comparison, the data for thick single crystal h-BN flake is shown together.
The curves are displaced vertically for clarity.
(b) 
$E_{\rm{1u}}$  phonon frequency $\omega_{0}$ against the h-BN thickness for the HT h-BN films ($1\sim40$ layers) and single crystal h-BN flakes ($>40$ layers).  
}
\label{fig:fig.3}
\end{figure}

High quality film growth often becomes challenging in the ultrathin, few-layer thickness region.
To investigate how the $E_{\rm{1u}}$ phonon change with thickness, we measured IR transmission for a series of HT h-BN samples with controlled thickness. 
Figure 3(a) shows $1-T/T_{0}$ of HT h-BN from $d=24$ layers to $d=1$ layer.
We used the relation 0.33~nm~=~1~layer to convert the unit of thickness from nm to the number of layers.
The characteristic phonon spectrum, prominent $E_{\rm{1u}}$ peak and absence of $A_{\rm{2u}}$ peak, persists for all samples, indicating the aligned growth of h-BN layer at few layer and monolayer film.
Interestingly the $E_{\rm{1u}}$ peak shifts to higher energy as the film becomes thinner. 
In Figure 3(b) we plot the phonon frequency ($\omega_{0}$) of h-BN film as function of h-BN thickness. 
$\omega_{0}$ measured for several exfoliated h-BN crystal flakes are shown together. 
For thick h-BN flakes ($d>40$ layers) $\omega_{0}$ is independent with $d$ and it agrees with the frequency of bulk h-BN crystal.
For HT h-BN films $\omega_{0}$ increases rapidly as $d$ decreases at $d<24$ layers. 
The steep increase of phonon frequency in the few-layer thickness is observed in other layered compounds such as topological insulator (Bi$_{2}$Se$_{3}$)\cite{aguilar2012terahertz} and transition metal dichalcogenide (MoS$_{2}$, MoTe$_{2}$)\cite{lee2010anomalous,yamamoto2014strong} suggesting that it is possibly an universal behavior of layered 2D materials.
The strong shift of phonon frequency shows that one can use $\omega_{0}$ to identify thickness of h-BN film as an alternative to AFM or TEM.

We discuss briefly about the phonon frequency of our samples. 
For HT h-BN film the phonon frequency agrees with that of the single crystal for both $A_{\rm{2u}}$ peak and $E_{\rm{1u}}$ peak. 
However, for LT h-BN film, the $A_{\rm{2u}}$ ($E_{\rm{1u}}$) phonon frequency is notable lower (higher) than HT h-BN. 
When the substrate temperature is lower than the optimal temperature during the growth (HT~=~1400~$\celsius$ for our case) 
the B and N atoms cannot crystallize into fully relaxed, equilibrium lattice due to insufficient migration energy, 
causing some strain or tension in the film. We propose that 
the lattice strain may be one possible reason for the phonon frequency shifts. Also LT h-BN film may contain atomic vacancy and defects,
which may also cause the phonon shifts.

\subsection{Graphene on h-BN}

\begin{table}[b]
\caption{\label{tab:table1}%
Dirac carrier parameters measured from the interband transition conductivity $\sigma_{1}$ of graphene on different substrates. 
$E_{\rm{F}}$ = Fermi energy, $n$= carrier density, and Temp$^{*}$ = effective temperature.
}
\begin{ruledtabular}
\begin{tabular}{cccc}
Substrate & $E_{\rm{F}}$ (meV) & $n$ (10$^{12}$~cm$^{-2}$) & Temp$^{*}$ (K) \\
\hline
HT h-BN (2.5~nm)	& 93	&  0.5 & 300 \\
LT h-BN (40~nm)		& 155	&  1.5 & 370 \\
SiO$_{2}$			& 214	&  2.9 & 400 \\
\end{tabular}
\end{ruledtabular}
\end{table}

\begin{figure}[t!]
\centering
\epsfig{file=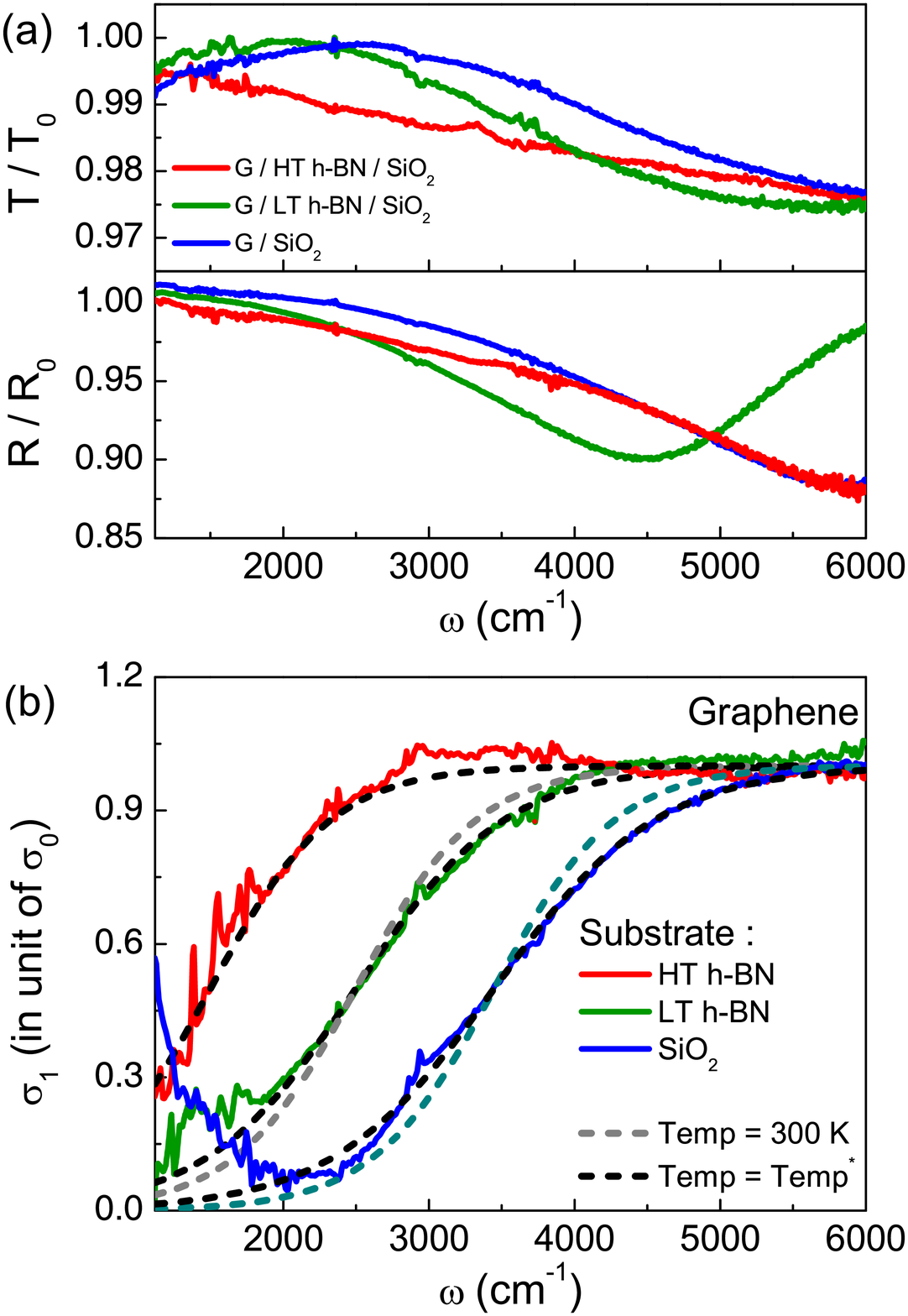,width=0.40\textwidth,angle=0,clip=}
\caption{
(a) 
Transmission and reflection of graphene deposited on substrate, G/substrate, for three different substrates, HT h-BN/SiO$_{2}$, LT h-BN/SiO$_{2}$, and SiO$_{2}$.
(b)
Optical conductivity $\sigma_{1}$ of graphene calculated from $T/T_{0}$ and $R/R_{0}$. 
Theoretical fits (see text) are shown by dashed curves.
}
\label{fig:fig.4}
\end{figure}

To study the role of h-BN thin film as substrate for large scale graphene we transfer monolayer CVD graphene sheet on three different substrates, HT h-BN(2.5~nm~thick)/SiO$_{2}$/Si, LT h-BN(40~nm~thick)/SiO$_{2}$/Si, and bare SiO$_{2}$/Si. 
Figure 4(a) shows the infrared transmission $T/T_{0}$ and reflection $R/R_{0}$ measured for the graphene(G)/substrate, where again the spectra $T$ and $R$ are normalized by the substrate $T_{0}$ and $R_{0}$. 
We extracted optical conductivity of graphene alone by applying the Kramers-Kronig constrained variational analysis to $T/T_{0}$ and $R/R_{0}$.\cite{kuzmenko2005kramers}  

Figure 4(b) shows optical conductivity $\sigma_{1}$ of graphene.
$\sigma_{1}$ represents interband transition $\sigma_{1}=\sigma_{0}(=\pi e^{2}/2h)$ of the Dirac carrier.\cite{kravets2010spectroscopic,chae2011excitonic}
At low $\omega$, $\sigma_{1}$ is suppressed because the transition becomes Pauli-forbidden.\cite{li2008dirac}
The $\sigma_{1}$ curve shifts to lower energy as substrate is changed in the order SiO$_{2}$ $\rightarrow$ LT h-BN $\rightarrow$ HT h-BN.
Theoretically the interband optical transition is expressed as 
\begin{equation}
\sigma_{1}=\frac{\sigma_{0}}{2} \left[ \tanh \left( \frac{\hbar\omega+2E_{\rm{F}}}{4k_{\rm{B}}T} \right)
+\tanh \left( \frac{\hbar\omega-2E_{\rm{F}}}{4k_{\rm{B}}T} \right)\right]
\end{equation} 
where $E_{\rm{F}}$ and $T$(=Temp) stand for the Fermi energy and sample temperature respectively.\cite{mak2008measurement} 
We fit $\sigma_{1}$ using Eq.(1). 
The fitting parameters are summarized in Table I. 
For graphene on SiO$_{2}$, we have $E_{\rm{F}}=214$~meV from the fit which, from the relation $E_{\rm{F}}=\hbar v_{\rm{F}} \sqrt{\pi n}$ and $v_{\rm{F}}=1.1\times10^{8}$~cm/s, shows that Dirac band is doped by carrier, known to be hole, with density $n=2.9\times10^{12}$~cm$^{-2}$.
The hole-doping of graphene is caused by various defect of SiO$_{2}$ such as oxygen vacancy, surface dangle bonds, and others.
For HT h-BN, we have significantly less values $E_{\rm{F}}=93$~meV and $n=5\times10^{11}$~cm$^{-2}$. 
The latter is only 17~\% of that for G/SiO$_{2}$, showing that residual carrier has strongly decreased.
It is interesting to compare the large scale CVD G/h-BN with single crystal G/h-BN.
For exfoliated graphene on exfoliated h-BN, transport measurements showed $n$ is as small as $1.5\times10^{11}$~cm$^{-2}$ at room-$T$ which is 10~\% of $n$ for single crystal G/SiO$_{2}$ and 5~\% of $n$ for CVD G/SiO$_{2}$.\cite{dean2010boron}
To gain better idea on the $n$'s, we convert $n$ to the familiar transport $I$-$V$ curve of graphene. 
In the $I$-$V$ curve charge neutral voltage $V_{\rm{CNP}}$ is determined by $n$ as $n=(\epsilon_{\rm{r}}\epsilon_{0}/ed)V_{\rm{CNP}}$ where $\epsilon_{\rm{r}}=3.9$ and $d=300$~nm are dielectric constant and thickness of SiO$_{2}$, respectively.
For G/SiO$_{2}$ sample we have $V_{\rm{CNP}}=40$~V. 
It is reduced to $V_{\rm{CNP}}=7$~V for G/HT h-BN and $V_{\rm{CNP}}=2$~V for the single crystal G/h-BN. 
This result shows that large scale graphene was brought closer to the graphene on single crystal h-BN by use of the CVD h-BN, and in fact is not far from $V_{\rm{CNP}}=0$~V, the ideal free standing state. 
For the G/LT h-BN, we have $E_{\rm{F}}=155$~meV and $n=1.5\times10^{12}$~cm$^{-2}$. 
Those values lie between SiO$_{2}$ and HT h-BN showing that LT h-BN is not as efficient as HT h-BN although it does reduce the residual carrier to some extent. 
High quality h-BN with well aligned BN planes is essential to achieve residual carrier-free graphene.

For Temp~=~0, Eq.(1) predicts that $\sigma_{1}$ changes suddenly from $\sigma_{1}=0$ to $\sigma_{1}=\sigma_{0}$ at $\hbar\omega=2E_{\rm{F}}$. 
At elevated temperature (Temp~$>$~0), the carrier is thermally populated in the Dirac band and the sharp transition of $\sigma_{1}$ becomes broadened. 
For G/HT h-BN, $\sigma_{1}$ data agrees well with Eq.(1) calculated with the measurement temperature Temp~=~300~K. 
However for G/LT h-BN and G/SiO$_{2}$ the $\sigma_{1}$ transition is broader than the theory curve. 
Here $\sigma_{1}$ can be fit only using higher Temp, Temp~=~370~K for LT h-BN and Temp~=~400~K for SiO$_{2}$ as shown by black dashed curve.
We define for convenience those Temp as effective temperature Temp$^{*}$.
Temp$^{*}$~$>$~300~K implies that there are extra source other than temperature that broadens the curve. 
Possibly the carrier is populated or redistributed in the Dirac band due to the defect-driven perturbation from the substrate, LT h-BN and SiO$_{2}$.  
This postulation is supported by that Temp$^{*}$ is correlated with $E_{\rm{F}}$, i.e, Temp$^{*}$ decreases along with the residual carrier reduction as the substrate quality is increased as SiO$_{2}$ $\rightarrow$ LT h-BN $\rightarrow$ HT h-BN. 

\section{Conclusion}

In conclusion we performed infrared measurement on large scale CVD h-BN thin film and CVD graphene/h-BN heterostructure.
For CVD h-BN sample, HT (Temp~=~1400~$\celsius$) -grown thin film shows only $E_{\rm{1u}}$ optical phonon like single crystal h-BN flake, demonstrating that 2D h-BN planes are highly aligned over the entire area of the film.   
On the other hand, LT (1000~$\celsius$) -grown film shows two phonon peaks, $E_{\rm{1u}}$ and $A_{\rm{2u}}$, due to tilting of the h-BN plane by $\theta=20\sim40\degree$. 
For G/HT h-BN samples, interband transition spectrum $\sigma_{1}$ shows that residual carrier density of graphene is strongly suppressed to $n=5\times10^{11}$~cm$^{-2}$, from $n=2.9\times10^{12}$~cm$^{-2}$ of G/SiO$_{2}$ and becomes comparable to $n$ of single crystal graphene on single crystal h-BN.
Also the interband transition width of $\sigma_{1}$ is reduced from Temp$^{*}$~$=$~400~K for G/SiO$_{2}$ to the theoretical value Temp$^{*}$~$=$~300~K by using the HT h-BN film as substrate.
The behaviors of $n$ and Temp$^{*}$ show that HT h-BN can decouple large scale graphene from SiO$_{2}$ leading it closer to ideal, undoped Dirac semi-metallic state.

\begin{acknowledgments}
This work was supported by Mid-career Researcher Program through National Research Foundation (NRF) grant funded by the MEST (NRF-2014R1A2A2A01003448)
and a grant from the Center for Advanced Soft Electronics under the Global Frontier Research Program through the NRF funded by the Ministry of Science, ICT and Future Planning (2011-0031630).
\end{acknowledgments}

%

\end{document}